\documentclass{elsart}

\usepackage{natbib}

 \usepackage{graphicx}

\usepackage{amssymb}

\begin{document}

\begin{frontmatter}



\title{Global architecture of metabolite distributions across species and its formation mechanisms}


\author[JST,Tokyo,corr]{Kazuhiro Takemoto}
\ead{takemoto@cb.k.u-tokyo.ac.jp}

\address[JST]{PRESTO, Japan Science and Technology Agency, Kawaguchi, Saitama 332-0012, Japan}
\address[Tokyo]{Department of Computational Biology, Graduate School of Frontier Sciences, University of Tokyo, Kashiwanoha 5-1-5, Kashiwa, Chiba 277-8561, Japan}



\corauth[corr]{Corresponding author.}

\begin{abstract}
Living organisms produce metabolites of many types via their metabolisms.
Especially, flavonoids, a kind of secondary metabolites, of plant species are interesting examples.
Since plant species are believed to have specific flavonoids with respect to diverse environment, elucidation of design principles of metabolite distributions across plant species is important to understand metabolite diversity and plant evolution.
In the previous work, we found heterogeneous connectivity in metabolite distributions, and proposed a simple model to explain a possible origin of heterogeneous connectivity.
In this paper, we show further structural properties in the metabolite distribution among families inspired by analogy with plant-animal mutualistic networks: nested structure and modular structure.
An earlier model represents that these structural properties in bipartite relationships are determined based on traits of elements and external factors. 
However, we find that the architecture of metabolite distributions is described by simple evolution processes without trait-based mechanisms by comparison between our model and the earlier model.
Our model can better predict nested structure and modular structure in addition to heterogeneous connectivity both qualitatively and quantitatively.
This finding implies an alternative possible origin of these structural properties, and suggests simpler formation mechanisms of metabolite distributions across plant species than expected.
\end{abstract}


\end{frontmatter}

\section{Introduction}
\label{sec:intro}
How do metabolites distribute among species?
Metabolite distributions across species or comprehensive species-metabolite relationships \citep{Shinbo2006} are important to understand design principles for metabolism in addition to metabolic networks \citep{Barabasi2004,Papp2009} because living organisms produce metabolic compounds of many types via their metabolisms, which adaptively shape-shift with changing environment in a long evolutionary history.
Especially, since living organisms have specific metabolite compositions due to metabolisms adaptively changing with respect to the environment, we can estimate environmental adaptation (adaptive evolution) using metabolite distributions.

Toward this end, we used flavonoids to investigate structures of metabolite distributions among plant species in the previous work \citep{Takemoto2009}.
Flavonoids are especially interesting examples when considering such metabolite distributions.
Plant species have secondary metabolites of many types including flavonoids, alkanoids, terpenoids, phenolics, and other compounds.
These metabolites are not essential for preserving life unlike basic metabolites such as bases, amino acids, and sugars; however, they play additional roles aiding survival in diverse environments.
Therefore, distributions of secondary metabolites are believed to be significantly different among species due to environmental adaptation, implying high species specificity \citep{Gershenzon1983}.
For this reason, secondary metabolites help us to understand adaptation and evolution.

Metabolite distributions are represented as bipartite networks (or graphs) in which nodes of two types correspond to plant species and flavonoids and links denote species-flavonoid relationships.
In the previous work, we found heterogeneous connectivity (degree distribution) in the flavonoid distributions: the number of flavonoids in a plant species and the number of plant species sharing a flavonoid follow power-law-like distributions.
Moreover, a bipartite network model was proposed by considering simple evolution processes in order to explain a possible origin of the heterogeneous connectivity.
We showed that the model is in good agreement with real data with analytical and numerical solutions.

Bipartite relationships such as the above metabolite distributions among species are observed in other fields. 
A good example is plant-animal mutualistic networks, which occupy an important place in theoretical ecology and are important to understand cooperation dynamics and biodiversity.
In networks of this type, we can also observe the heterogeneous connectivity, or diversified patterns of interaction among species (both plants and animals) \citep{Jordano2003,Saavedra2009}.
In addition to this, non-random structural patterns such as nested structure \citep{Bascompte2003} and modular structure \citep{Olesen2007} were found in the mutualistic networks.
The nested structure means that animals (pollinators or seed dispersers) of a certain plant form a subset of those of another plant in a hierarchical fashion.
Such non-random patterns often strongly control dynamics of ecological systems \citep{Bastolla2009}.
The modular structure represents that the subsets of species (modules), in which species are strongly interconnected, are weakly connected.
Thus, this structural property helps to understand coevolution of two objects (i.e. plants and animals).

To reveal the origin of non-random patterns, \citet{Saavedra2009} proposed the bipartite cooperation (BC) model inspired by food-web models based on traits of species and external factors [reviewed in \citep{Stouffer2005}].
Although it agrees well with real plant-animal mutualistic networks, the BC model is a non-growth model in which the number of species (plants and pollinators) is fixed (i.e. this model is not an evolutionary model).
The structure of model-generated networks is determined by intrinsic parameters of three types drawn from exponential or beta distributions: foraging traits (e.g. efficiency and morphology), reward traits (e.g. quantity and quality) and external factors such as environmental context (e.g. geographic and temporal variation).
According to the above mechanism, the model network is generated with three observable parameters: the numbers of nodes of two types (e.g. the number of plants and the number of animals) and the number of interactions, and it is in good agreement with real mutualistic networks.
Furthermore, these structural properties are also observed in manufacturer-contractor interactions, and the BC model could reproduce them.
Therefore, it is believed that the BC model is a general model for bipartite relationships.

Taken together, several striking structural properties (i.e. heterogeneous connectivity, nested structure, and modular structure) are widely observed in bipartite networks, and there are two models to explain design principles for such bipartite networks: trait-based (non-evolutionary) model and evolutionary model.
Due to this, we had the following questions:
(i) Do metabolite distributions additionally show nested and modular structures in analogy with ecological networks and organizational networks?
(ii) Can our model \citep{Takemoto2009} reproduce nested and modular structures in addition to heterogeneous connectivity?
In other words, are these structural properties acquired in evolutionary history?
\citet{Rezende2007-1,Rezende2007-2} suggest that the structure of mutualism between plants and animals is affected by not only traits of species and external factors but also evolution processes.
Thus, it is expected that our model (i.e. evolutionary model) also can reproduce such non-random patterns.
(iii) Which is appropriate to our model (evolution process) and the BC model (trait-based mechanism) to describe the formation of metabolite distributions?


In this paper, we represent that metabolite distributions across species have nested structure and modular structure, and numerically investigate whether our model and the BC model can reproduce such non-random structures or not.
Furthermore, the prediction of network connectivity (degree distribution) is also evaluated between our model and the BC model.
From these results, we show that formation mechanisms of metabolite distributions across plant species are governed by simple evolution processes rather than traits of metabolites and plant species and external factors.

\section{Methods}
\subsection{Dataset}
We utilized the data in \citep{Takemoto2009} in which a total of 14,378 species-flavonoid pairs were downloaded from Metabolomics.JP \citep{Arita2008} (http://metabolomics.jp/wiki/Category:FL).
In this dataset, there are 4725 species and 6846 identified flavonoids.
The taxonomy (family) of a species was assigned according to The Taxonomicon (http://taxonomicon.taxonomy.nl).
The six largest families in terms of the number of reported flavonoids are considered: Fabaceae (bean family), Asteraceae (composite family), Lamiaceae (Japanese basil family), Rutaceae (citrus family), Moraceae (mulberry family), and Rosaceae (rose family).
We extracted species-flavonoid pairs from the dataset based on these six families, and constructed the metabolite distribution of each family using bipartite networks.


\subsection{The Model}
\begin{figure}[tbhp]
\begin{center}
	\includegraphics{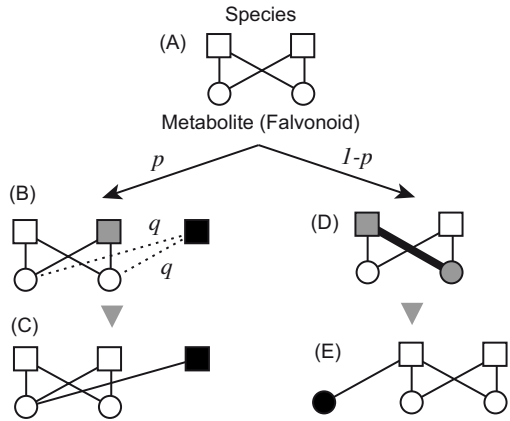}  
	\caption{
	Schematic diagram of our model.
	Squares and circles mean plant species and flavonoids, respectively.
	(A) An initial metabolite distributions represented as bipartite networks.
	(B) and (C) The addition of a new species and their interactions.
	The gray box represents a randomly selected species.
	The filled box indicates a new species resulting due to duplication of the selected species.
	The dashed lines are possible interactions between the new species and metabolites.
	(D) and (E) The addition of a new metabolite and their interaction.
	The thick link between gray nodes corresponds to a randomly selected species-metabolite pair.
	The filled circle means a new metabolite.
	}
	\label{fig:model}
\end{center}
\end{figure}

We here review our model proposed in \citep{Takemoto2009}.
In this model, a small initial metabolite distributions (Fig. \ref{fig:model} A) are first prepared, and it evolves according to two simple evolutionary mechanisms as follows:

(i) Metabolite compositions of new species are inherited from those of existing (ancestral) species.
We assume that new species emerge due to mutation of ancestral species.
In our model, this event occurs with the probability $p$ at time $t$, and new species are born from randomly selected existing species.
Flavonoid compositions of new species are inherited from that of ancestral species because new species are similar to the ancestral species due to mutation (Fig. \ref{fig:model} B).
By considering divergence, however, we model that each flavonoid is inherited from that of ancestral species with the probability $q$ (Fig. \ref{fig:model} C). 
Independently of our model, in addition, a bipartite network model generated based on the above inheritance (or copy) mechanism was proposed in \citep{Nacher2009} to describe evolution of protein domain networks around the same time.

(ii) New flavonoids are generated by variation of existing flavonoids.
In evolutionary history, living organisms accordingly obtain new metabolic enzymes via gene duplications \citep{Diaz-Mejia2007} and horizontal gene transfers \citep{Pal2005}, and the metabolic enzymes synthesize new metabolites through modification of existing flavonoids with substituent groups and functional groups.
We model that this event occurs with the probability $1-p$ at time $t$ and a species-flavonoid pair is selected at random (Fig. \ref{fig:model} D), and its species obtains a new flavonoid (Fig. \ref{fig:model} E).

Our model have two parameters $p$ and $q$.
We can generate the model network through the estimation of the parameters $p$ and $q$ using observable parameters of real metabolite distributions: the number of plant species $S$, the number of metabolites (flavonoid) $F$, and the number of interactions $L$. 

The parameter $p$ is estimated as
\begin{equation}
p=\frac{S}{S+F}
\label{eq:p}
\end{equation}
because $S=pt$ and $F=(1-p)t$ in our model.

To obtain the parameter $q$, we need to consider the time evolution of $L$.
This is derived as $L\approx (1-p)t/(1-q)$ \citep{Takemoto2009}. 
Since $F=(1-p)t$, as above, the parameter $q$ is estimated as
\begin{equation}
q=1-F/L.
\label{eq:q}
\end{equation}

Using Eq.s (\ref{eq:p}) and (\ref{eq:q}), we estimated the parameters $p$ and $q$ from real data, and generated corresponding model networks for comparison with real ones.

\section{Result}
We first investigated the nestedness and the modularity of metabolite (flavonoid) distributions.
To measure the degrees of nested structure and modular structure (i.e. nestedness $N$ and modularity $Q$) of metabolite distributions, we utilized the BINMATNEST program \citep{Girones2006} and the optimization algorithm proposed in \citep{Guimera2005}, respectively.
The nestedness $N$ ranges from perfect non-nestedness ($N=0$) to perfect nestedness ($N=1$), and the high modularity $Q$ means a strong modular structure. 
We also calculated $N$ and $Q$ from randomized networks generated by the null model 2 in \citep{Bascompte2003} in order to show statistical significance of the structural properties.
The statistically significance is suitably evaluated because the null model 2 generates randomized networks without bias of heterogeneous connectivity. 

\begin{figure}[tbhp]
\begin{center}
	\includegraphics{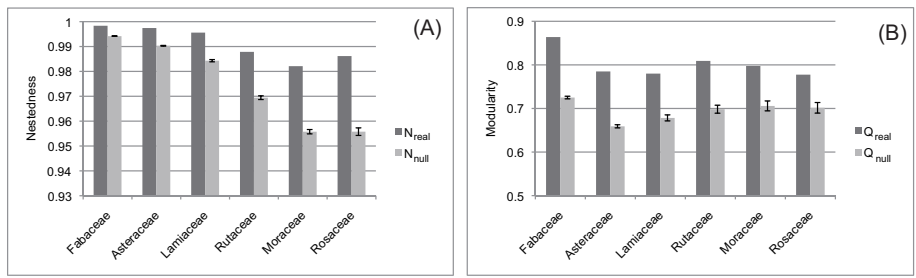}  
	\caption{
	Significant high nestedness $N$ (A) and modularity $Q$ (B) in metabolite distributions across plant species.
	The dark gray bars and the light gray bars correspond to real values and the null model, respectively.
	$N$ and $Q$ obtained from the null model are averaged over 100 realizations.
	All $P$-values for the difference are lower than 0.0001.
	The $P$-value is derived using the $Z$-score defined as $(x_{real}-\bar{x}_{null})/SE_{null}$, where $x_{real}$ corresponds to real values (nestedness or modularity).
	$\bar{x}_{null}$ and $SE_{null}$ are the average of values from the null model and its standard error, respectively.
	}
	\label{fig:significant_NQ}
\end{center}
\end{figure}

Figure \ref{fig:significant_NQ} shows the comparison of nestedness $N$ and modularity $Q$ between real data and the null model for each family.
As shown in this figure, $N$ and $Q$ of real data are significantly larger than that of the null model, indicating that metabolite distributions also show nested structure and modular structure in addition to heterogeneous connectivity as ecological networks and organizational networks.
In addition, the nestedness and the modularity are different structural properties because of no correlation between them (Pearson correlation coefficient $r=0.345$ with $P$-value $p=0.503$).
The above result means that metabolites in a plant species is a subset of that in other plant species, and plant species are divided into several clusters based on their metabolite compositions.

\begin{figure}[tbhp]
\begin{center}
	\includegraphics{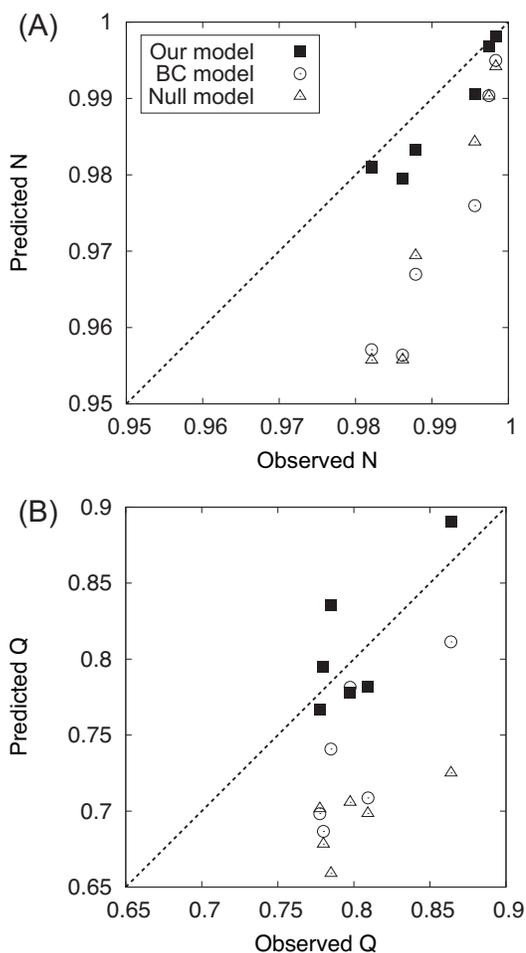}  
	\caption{
	Comparison of nestedness $N$ (A) and modularity $Q$ (B) between models and real data.
	The dashed line represents the perfect agreement between predicted values ($N$ or $Q$) and observed ones.
	The nestedness and the modularity from models are averaged over 100 realizations.
	}
	\label{fig:nestedness}
\end{center}
\end{figure}

Next, the prediction of nestedness $N$ and modularity $Q$ by our model and the BC model was mentioned.
Figure \ref{fig:nestedness} shows the comparison of $N$ and $Q$ between models and real data.
For comparison, we also computed $N$ and $Q$ calculated from the null model.

We evaluated the prediction accuracy of our model and the BC model using the Pearson correlation coefficient (CC) and the root mean square error (RMSE) between predicted values $x_i$ and observed values $y_i$, defined as
\[
\mathrm{RMSE}=\sqrt{\frac{1}{n}\sum_{i=1}^n(x_i-y_i)^2},
\]
where $n$ is the number of samples (i.e. the number of families $n=6$).
The CC and the RMSE represent the degrees of agreement and error between observed values and predicted ones, respectively.
Our model showed the higher CCs and the lower RSMEs (see Table \ref{table:compari_nest}), indicating that our model has the higher prediction accuracy than the BC model. 

\begin{table}[tbhp]
\caption{Prediction accuracy for nestedness and modularity: the correlation coefficient (CC) and the root mean square error (RMSE). The emphasized values correspond to the best accuracy.}
\label{table:compari_nest}
\begin{center}
\begin{tabular}{l|cc|cc}
\hline
\hline
& \multicolumn{2}{c|}{Nestedness} & \multicolumn{2}{c}{Modularity} \\
& CC & RMSE & CC & RMSE \\
\hline
Our model & {\bf 0.949} & {\bf 0.0039} & {\bf 0.770} & {\bf 0.0282} \\
BC model & 0.946 & 0.0200 & 0.767 & 0.0708 \\
\hline
\hline
\end{tabular}
\end{center}
\end{table}



\begin{figure}[tbhp]
\begin{center}
	\includegraphics{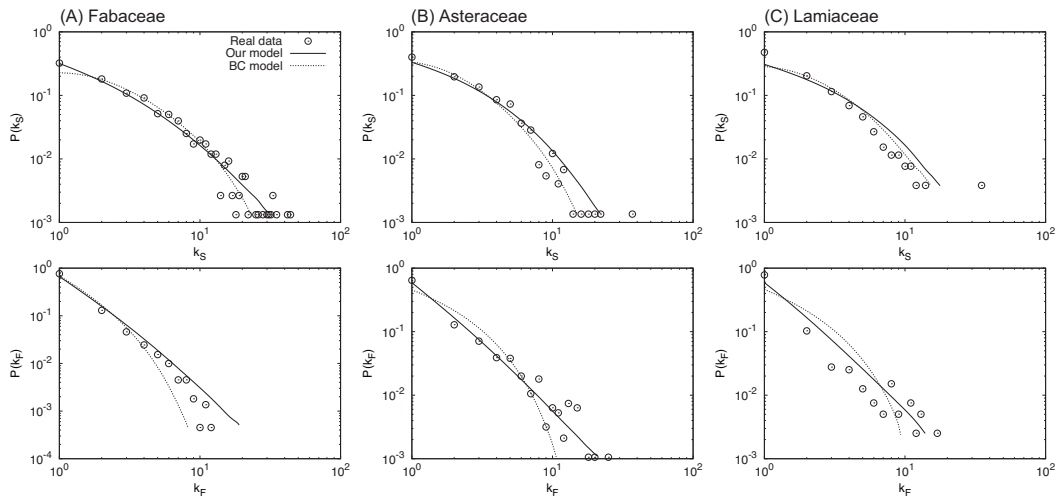}  
	\caption{
	Heterogeneous degree distributions in metabolite distributions for plant species (top column) and metabolites (flavonoids) (bottom column).
	The circles and the lines represent real data and models, respectively.
	The degree distributions of models are averaged over 100 realizations.}
	\label{fig:degree}
\end{center}
\end{figure}

We finally considered the frequency of the number of interactions per nodes (degree distribution).
Figure \ref{fig:degree} shows the degree distributions of metabolite distributions (symbols) and the models (lines).
The degree distributions of the only three metabolite distributions as representative examples due to the space limitation.
We could observe degree distributions of two types [$P(k_S)$ and $P(k_F)$, where $k_S$ and $k_F$ denote the degrees of nodes corresponding to plant species and metabolites (flavonoids), respectively] because metabolite distributions are represented as bipartite graphs.
In the both cases, the degree distributions follow a power law with an exponential truncation, and model-generated degree distributions are in good agreement with real ones.
However, the BC model seems to have bad predictions in the case of $P(k_F)$.

To quantitatively verify goodness of fits between models and real data, we calculated the tail-weighted Kolmogorov-Smirnov (wKS) statistics (distance) \citep{Clauset2009}, defined as
\[
\mathrm{wKS}=\max_{x}\frac{|R(x)-M(x)|}{\sqrt{M(x)[1-M(x)]}},
\]
between empirical distributions $R(x)$ and predicted distributions $M(x)$ for species nodes (wKS$_S$) and metabolite (flavonoid) nodes (wKS$_F$).

\begin{figure}[tbhp]
\begin{center}
	\includegraphics{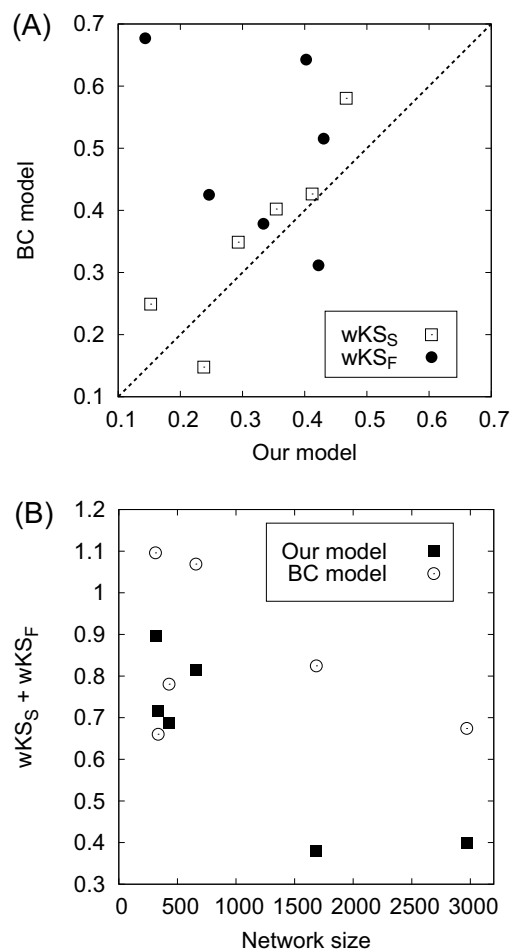}  
	\caption{
	(A) Comparison of the tail-weighted Kolmogorov-Smirnov statistics (distances) between our model and the BC model.
	The dashed line corresponds to the same prediction accuracy between our model and the BC model.
	(B) Correlation between network size and wKS distance.
	}
	\label{fig:ks_distance}
\end{center}
\end{figure}

Figure \ref{fig:ks_distance} A shows the comparison of wKS distances between our model and the BC model.
In the case of $P(k_S)$ (i.e. wKS$_S$), the prediction accuracy (wKS distance) is almost similar between our model and the BC model.
In the case of $P(k_F)$, however, we can find the critical difference of the prediction between our model and the BC model.
Our model could more highly predict $P(k_F)$ than the BC model.

Figure \ref{fig:ks_distance} B shows the correlation between the network size (i.e. $S+F$) and prediction accuracy, defined as wKS$_S+$wKS$_F$.
As shown in this figure, the prediction accuracy of our model tends to decrease with the network size ($r=-0.850$ with $p=0.032$), suggesting better predictions of our model for degree distributions in the case of larger networks.
However, there is no correlation between the prediction accuracy and network size in the case of the BC model ($r=-0.436$ with $p=0.387$).

\section{Discussions}
In summary, metabolite distributions across plant species also show nested structure and modular structure in addition to heterogeneous connectivity in analogy with plant-animal mutualistic networks and organizational networks, suggesting that such structural properties are universal among bipartite networks in wide-ranging fields.
Moreover, we found that our model can also reproduce these structural properties in addition to the BC model, indicating an alternative way to obtain these structural properties.
In other words, we showed that there are two different ways to obtain the structural properties: the trait-based way (the BC model) and the evolutionary way (our model).

Either one of these two way (i.e. the BC model or our model) might become significant due to the types of bipartite relationship and observation condition. 
In particular, metabolite distributions might be different from ecological networks and organizational networks in perspective of design principles despite the same structural properties.
As above, we showed that our model could better reproduce such structural properties of metabolite distributions than the BC model. 
This finding implies that these structural properties of metabolite distributions are acquired through evolution processes, considered in our model, rather than trait-based mechanisms (i.e. the BC model), believed to be a general formation mechanism of bipartite networks.

Compared to ecological networks and organizational networks, metabolite distributions might be hardly influenced by traits of elements (i.e. plant species and metabolites) and external factors.
This might because we can observe comprehensive species-metabolite relationships.
In the case of ecological networks and organizational networks, such observations might be difficult.
For example, we assume that a plant species can interact to pollinators A, B, C and D in plant-animal mutualistic networks.
However, these interactions are limited because of several conditions such as geography and pollinators' properties (e.g. environmental fitness).
Supposing that the pollinator A only lives in area I, and the rest (i.e. B--D) is in area II due to such conditions, we can find the different mutualistic networks between areas I and II.
Because of such restraints, ecological networks might be different from metabolite distributions.
However, we speculate that our model can be also applied to ecological networks of this type if plant-animal relationships are comprehensively obtained under ideal conditions (e.g. environmentally homogeneous islands).
In fact, our model could reproduce the structure of plant-animal mutualistic networks in a limited way \citep{Takemoto2009-2}.

When we consider the global tendency of bipartite relationships such as nested structure, modular structure, and heterogeneous connectivity, our model can explain its origin more simply than the BC model.
This is an advantage of our model.
In the case of the BC model, the formation mechanisms are relatively complicated because we need to consider the interaction rule based on traits between elements and external factors.
However, we believe that elements' traits and external factors are important.
Especially such factors might play crucial roles for the formation of local interaction patterns.

Using our model, the formation mechanisms of the structural properties in metabolite distributions are described as follows.

The nested structure means that a plant's flavonoid composition is a subset of other plants' flavonoid compositions, and its origin is explained using our model as follows.
In our model, metabolites of a new plant are inherited from those of an ancestral plant because these plants tend to be similar due to mutation.
However, new plants obtain the part of metabolites by considering divergence (elimination of interactions).
As a result, metabolites of an offspring plant become a subset of those of their parent plant, and produce nested structure.

The modular structure implies that plant species are divided into several clusters in which they are strongly interconnected through common metabolites and these clusters interact loosely.
In short, the modular structure is obtained by strong interconnections in clusters and weak interactions among these clusters.
Emergence of weak and strong interactions is also described by inherence and divergence of metabolite compositions.
As above, metabolite compositions are inherited from ancestral species in our model.
Then, new species and ancestral species are connected because of common metabolites, and interactions of this type correspond to strong interconnections.
Due to divergence, on the other hand, new species indirectly connect to the other species via metabolites of ancestral species that were not inherited by new species, and this results weak interactions.

Regarding the origin of heterogeneous connectivity, we have already discussed in \citep{Takemoto2009}.
Easily speaking, the duplication mechanism and the randomly selection of species-flavonoid pairs result preferential attachments (`rich-gets-richer' mechanisms) because nodes with many neighbors tend to obtain more neighbors when considering such mechanisms.
These are strongly related to the duplication-divergence model \citep{Vazquez2003} and the Dorogovtsev-Mendes-Samukhin model \citep{Dorogovtsev2001}, respectively.

In the case of metabolite distributions, as above, we believe that nested structure, modular structure, and heterogeneous connectivity are dominantly acquired in evolutionary history.
Thus, these structural properties might provide novel classification schemes of plant species based on metabolite compositions such as chemotaxonomy.
For example, we might be able to extract hierarchical organization of plant species based on their metabolite compositions from nested structure.
Moreover, modular structure might reveal classified characteristic species-metabolite relationships, and heterogeneous connectivity helps to find useful metabolites (i.e. hub metabolites) for taxonomic classification and characterization of plant species at higher levels (e.g. family and order).
As a result, these structural properties might provide insights into metabolite diversity and plant evolution.

For simplicity, we did not consider a number of important evolution processes (especially deletions of nodes and interactions) at present.
In particular, the degree distributions may become different due to such extinctions \citep{Enemark2007,Deng2007}.
However, such mechanisms might contribute only negligible effects on the above structural properties (the grobal tendency) according to our result.
This might be because such mechanisms tend to be nonessential in plant evolution.
In plant species, genome doubling (polyploidity) is a major driving force for increasing genome size and the number of genes \citep{Adams2005}.
Duplicated genes typically diversify in their function, and some acquire the ability to synthesize new compounds.
Indeed, plants acquire metabolites of many types (mostly secondary metabolites) \citep{DeLuca2000}, compared to a few thousand primary metabolites in higher animals.
The population of flavonoids, a type of secondary metabolites, is therefore expected to increase, indicating that we can roughly dismiss the effect of node losses when we consider the global tendency of metabolite distributions.
However, this does not mean that the deletions of nodes and interactions are unnecessary.
Such evolutionary mechanisms might play important roles to determine partial (or local) interaction patterns of bipartite relationships.
Thus, we need to focus on such evolution processes in the future to fully understand the formation of metabolite distributions across species.

\section*{Acknowledgment}
This work was supported by a PRESTO program of the Japan Science and Technology Agency.

\end{document}